\documentclass[journal]{IEEEtran}

\usepackage{graphicx}
\usepackage[nolist]{acronym}
\usepackage{amsmath} 
\usepackage{color,ulem}
\newacro{AI}[AI]{Artificial Intelligence}
\newacro{DL}[DL]{Deep Learning}
\newacro{SED}[SED]{Sound Event Detection}
\newacro{KD}[KD]{Knowledge Distillation}
\newacro{IoT}[IoT]{Internet of Things}
\newacro{IPS}[IPS]{Instructions Per Second}
\newacro{CNN}[CNN]{Convolutional Neural Network}
\newacro{GRU}[GRU]{Gated Recurrent Unit}
\newacro{RNN}[RNN]{Recurrent Neural Network}
\newacro{MIPS}[MIPS]{Million Instructions Per Second}
\newacro{SIMD}[SIMD]{Single Instruction Multiple Data}
\newacro{MCU}[MCU]{MicroController Unit}
\newacro{DSP}[DSP]{Digita Signal Processing}
\newacro{pdf}[pdf]{Probability Density Function}
\newacro{MSE}[MSE]{Mean Square Error}
\newacro{SQNR}[SQNR]{Signal to Quantization-Noise Ratio }
\newacro{MOPS}[MOPS]{Million Operation Per Second}
\newacro{MAC}[MAC]{Multiply-ACcumulate}
\usepackage[binary-units=true]{siunitx}  
\usepackage{textcomp}

\def\x{{\mathbf x}}
\def\X{{\mathbf X}}
\def\y{{\mathbf y}}
\def\Y{{\mathbf Y}}

\def\ti{\texttildelow{}}
\def\L{{\cal L}}

\begin{document}

\title{Compact recurrent neural networks for \\ acoustic event detection on low-energy low-complexity  platforms}


\author{\IEEEauthorblockN{Gianmarco Cerutti$^1$, Rahul Prasad$^2$, Alessio Brutti$^1$, Elisabetta Farella$^1$}\\
\IEEEauthorblockA{
\textit{$^{1}$ ICT-irst Fondazione Bruno Kessler}, Trento, Italy \\
\textit{$^2$School of Information Science, Manipal Academy of Higher Education, Manipal}, Karnataka, India\\ 
\{gcerutti, brutti, efarella\}@fbk.eu, prassd25@gmail.com}
}

%


\maketitle

\begin{abstract}
Outdoor acoustic event detection is an exciting research field but challenged by the need for complex algorithms and deep learning techniques, typically requiring many computational, memory, and energy resources. These challenges discourage IoT implementations, where an efficient use of resources is required. However, current embedded technologies and microcontrollers have increased their capabilities without penalizing energy efficiency. This paper addresses the application of sound event detection at the very edge, by optimizing deep learning techniques on resource-constrained embedded platforms for the IoT. 
The contribution is two-fold: firstly, a two-stage student-teacher approach is presented to make state-of-the-art neural networks for sound event detection fit on current microcontrollers; secondly, we test our approach on an ARM Cortex M4, particularly focusing on issues related to 8-bits quantization. Our embedded implementation can achieve 68\% accuracy in recognition on Urbansound8k, not far from state-of-the-art performance, with an inference time of 125 ms for each second of the audio stream, and power consumption of 5.5 mW in just 34.3 kB of RAM.
\end{abstract}

\section{Introduction}
\label{sec:intro}
\ac{IoT} applications require a large number of heterogeneous devices to be distributed in a certain environment. Each of them can potentially generate a large amount of data  
to be sent via wireless transmission, affecting the energy autonomy and lifetime of devices. In addition, privacy issues increase. One successful approach is distributing the computation at the edge, i.e. performing local pre-processing, but also advanced processing (e.g. machine learning, classification), directly on the wireless node, at "the thing" level~\cite{Conti_TCAS2017, Rusci_IoTJ2017}. Thanks to recent improvements in embedded technology, computationally powerful microcontrollers with consumption in the range of mW enables \ac{AI} at the edge. This reduces the amount of data transmitted, avoiding flooding an enormous quantity of raw data at the cloud level, and the related power consumption. 

\ac{SED} is an example of \ac{IoT} application where this approach can make a difference. In fact, since we are interested in events and not in raw data, a near-sensor processing approach is opportune. 
\ac{SED}, as well as acoustic scene recognition, can benefit from understanding events locally where they happen both in terms of privacy \cite{GarciaLopez:2015} and reaction time, which can be kept in the range of \si{\milli\second}. More than this, device lifetime can be guaranteed up to several years of operation, when energy harvesting is applied and the transmission is limited to few bytes. However, \ac{SED} is a rather challenging task, especially when applied in outdoor contexts. After some pioneering efforts~\cite{Temko-2007, Zhuang20101AED, Mesaros2010}, which have not led to established solutions, recent progresses in deep learning and the release of sound event datasets and challenges like \textit{UrbanSound8K} \cite{salamon2014dataset}, AudioSet \cite{audioset}, ESC50 \cite{piczak2015esc} and DCASE \cite{Mesaros2016_EUSIPCO, DCASE2017challenge} have reawakened interest in these applications, considerably improving the performance and paving the way to further developments.
Nevertheless, advances in terms of accuracy and robustness of current acoustic event detection algorithms are achieved by using large neural networks, which are increasingly hungry in terms of computational power and memory. This prevents the development of applications for distributed monitoring in public spaces, which require a pervasive network of energy neutral devices composed of cheap, low-power, low-complexity platforms. An attractive way to reduce the network complexity while preserving as much as possible its generalization capabilities is through \ac{KD}~\cite{hinton2015distilling}. Taking advantage of the redundancy that characterizes large networks~\cite{LeCun_1990}, \ac{KD} allows training small networks capable of mimicking large ones.

\begin{figure}[t]
  \centering
  \includegraphics[width=\columnwidth]{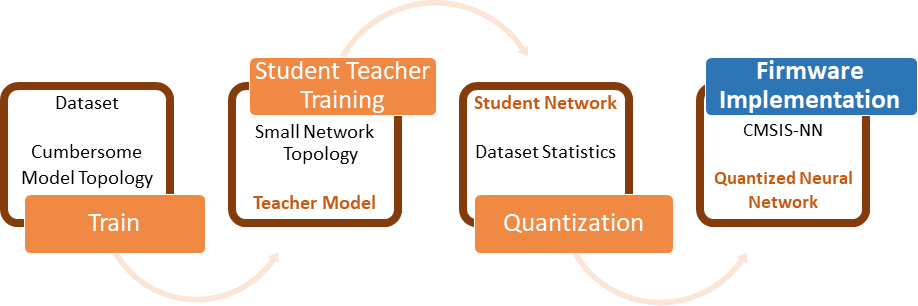}
  \caption{From state-of-the-art to \ac{IoT}: Starting from a sound event detection dataset and a state-of-the-art network topology that performs well on the problem, we apply the student-teacher approach to train a much smaller network that would fit on \ac{IoT} devices. Finally, we quantize the compact network and write the firmware for low-energy platforms.}
  \label{fig:flow}
  \vspace{-5pt}
\end{figure}

Going from state-of-the-art neural models to actual implementation on an IoT device involves multiple stages, as depicted in Fig~\ref{fig:flow}.
 In our previous publication~\cite{cerutti2019IS}, we presented a~\ac{KD} approach to compress a \ac{SED} classifier composed of the publicly available VGGish feature extractor~\cite{hershey2017cnn} and a recurrent classifier. Differently from common applications of~\ac{KD}, aimed at improving performance or at achieving limited reductions of the model dimensions, we obtained very high compression factors, reducing the network size from approximately 70 million parameters to nearly 20 thousand. These results paved the way to the effective use of deep neural networks on a low-power microcontroller to enable~\ac{SED}.

In this paper, we focus in particular on i) a preliminary analysis of the computational and memory requirements to understand what kind of models can be afforded by a given class of microcontrollers; ii) the quantization of the network parameters and the activations, presenting two different strategies to select for each layer the best fixed-point representation; iii) 
an implementation of the reduced network on a microcontroller with resources typical of an IoT end-node, building upon the network reduction strategies presented in~\cite{cerutti2019IS} and iv) the  evaluation of the accuracy of the actual implementation. 
In addition, we present an improvement of the~\ac{KD} approach presented in~\cite{cerutti2019IS}. Distillation is performed in two stages where adaptation of the VGGish pre-trained feature extraction to the in-domain data is separated from the actual parameter distillation, leading to a further improvement of the classification accuracy.



\section{Related Works} \label{sec:relatedwork}
Nowadays, deep neural networks are state-of-the-art for many classification tasks, like image classification and speech recognition. The trend is to use networks of continuously increasing size and complexity because they generalize better than shallower ones~\cite{canziani2016analysis}. As a consequence, the most recent and advanced solutions may be not practical if limited computational resources are available, as in the case of IoT application contexts.  
In fact, \ac{IoT} nodes can be used as simple sensing devices, transmitting all the information directly to the cloud (\cite{kelly2013towards, piyare2017plug}). Nevertheless, this transmission is usually expensive from an energy point of view; thus, processing data locally on the node can be preferred, especially when the throughput is considerably high (e.g., audio, inertial movement unit, video). Therefore, in those IoT-related scenarios, reducing the network complexity and preserving as much as possible of its generalization capability is of significant interest.
%
%
Fortunately, technology comes in to help, since the microcontrollers available nowadays are low-cost, energy-efficient processing units with average computation capability that allows non-trivial processing on board. Nevertheless, these systems still have some severe limitations, for example, in terms of memory limited to up to hundreds of \si{\kilo\byte}). 

Enabling advanced machine learning on \ac{IoT} nodes is, therefore, of great interest and is becoming an attractive research topic for a variety of digital signal processing applications.


One way to achieve edge deep learning is by employing non-commercial platforms optimized for neural networks. As an example, the authors of~\cite{palossi2018ultra} use a dedicated processing platform with state-of-the-art energy efficiency for an ultra-low-power deep-learning-powered autonomous nano-drone. In this way, no particular efforts are required in the network design thanks to the capabilities of the device. Processing time and memory footprint are further reduced by properly quantizing the network. Quantization is also relevant in \cite{cerutti2019convolutional} where, in combination with the use of the CMSIS-NN library, an embedded-C framework for neural network developed for Cortex M4-M7 based microcontrollers \cite{lai2018cmsis}, it allows a rapid and low-power classification of thermal images. However, the neural architecture is rather small (three-layer \ac{CNN}), and the resolution of the thermal images is shallow (8x8).

In the previous examples, either the device has adequate resources or the feature dimensionality is very small; thus, a deep learning algorithm can run on the embedded platform. The improvement obtained via quantization is somehow "imposed" by the fixed point representation typical of an efficient microcontroller. However, for other classification tasks, in particular those involving the processing of audio streams, the approaches presented above are not practicable. The reason is that extensive neural networks, as well as feature vectors, are employed in state-of-the-art solutions.


Keyword spotting is another field of interest that requires always-on smart devices near-to-the speaker and thus calls for energy-efficient embedded systems with on-board recognition capabilities. A common strategy is to design and implement small networks, expressly fitted to the hardware capacity, that can be trained from scratch and implemented on low-power low-cost microcontrollers. \cite{Sainat2015cnnkws} shows the superiority of~\ac{CNN} compared to fully connected deep neural networks in terms of performance, number of parameters and operations. Tang et al. implemented a set of residual neural networks with specific compact structures, focused on reducing the overall number of parameters and operations  \cite{Tang2018residual}. 
Zhang et al. implemented a keyword spotter on a commercial microcontroller using fixed-point quantization and a CMSIS-NN implementation~\cite{zhang2017hello}, obtaining very short inference times.

A completely different strategy is to compress an existing model, generating a new network with a smaller memory footprint, but that effectively mimics the original one. In literature, several approaches exist to reduce the number of parameters of a neural network. Network pruning \cite{hassibi1993optimal} aims at detecting and removing unimportant weights from a trained network until a given stop condition is reached. Matrix decomposition uses a compact format to represent the dense weight matrix of the fully-connected layers using few parameters, preserving the expressive power of the layer \cite{Novikov2015}. Matrix/tensor factorization~\cite{Denton_2014}, that exploits the linear structure of networks~\cite{Denil_2013}, and vector quantization of weights~\cite{Wu_2015} are other strategies to reduce the network memory size. These methods reduce the amount of memory needed to accommodate the network (e.g., sharing the weights). However, they keep the same architecture, therefore requiring the same buffers (RAM) and throughput. Besides, network pruning requires a manual setup of sensitivity for each layer and fine-tuning of the parameters.

A further way to achieve model compression is to reduce the weight representation to very few bits. For example, BinaryConnect~\cite{Courbariaux_2015} and the related Binary Weight Net~\cite{Rastegari_2016} represent weights with only 2 bits. If properly trained, the quantized networks can achieve performance close to the floating-point original models also on complex classification tasks~\cite{Leng_2018}. However, the memory and computational cost reductions are not sufficient for implementation on IoT devices where few KB are available (going from 32 to 2 bits results in a compression factor of 16 in the memory footprint). Experiments in~\cite{Leng_2018}, in fact, do not address the low-cost low-power devices we are targeting here. Additionally, non-conventional frameworks are needed to train the quantized network.
%
%

An attractive approach is to compress networks into a different and simpler architecture via \ac{KD} \cite{buciluǎ2006model}. This approach is also referred to as Student-Teacher because the smaller network (student) is trained to mimic the output of the larger one (teacher) \cite{shen2016teacher}.
The underlying idea is that the output of the neural network (soft labels) is more abundant in information than the hard labels and makes the training easier \cite{hinton2015distilling}. An example of network compression related to~\ac{SED} is \cite{kumariedgel}. Starting from the L$^3$ network for embedding extraction trained through self-supervised learning of audio-visual correspondence in videos~\cite{cramer2019look}, Kumari et al. compress this network targeting small edge devices, such as "motes" that use microcontrollers and achieve long-life self-powered operation. The work investigates the merging of different compression techniques (pruning,~\ac{KD}) and highlights the increase of performance using fine-tuning after compression.

Our proposed approach differentiates itself from those available in literature in multiple directions since it attempts to pull together the benefits of the methods reported above. We start from a large model and compress it. However, instead of just focusing on pruning or weight sharing, which provides limited memory reduction without decreasing the processing time, we design a minimal target network and use~\ac{KD} to train it (instead of training from scratch as in~\cite{Sainat2015cnnkws,Tang2018residual,zhang2017hello}). Note that distillation is typically employed to obtain limited network reductions while, in this paper, we target extremely high compression factors. On top of this heavy compression, we apply a stochastic weight and buffer quantization without the need for retraining the network.

The overview provided in this section shows the scientific interest in bringing intelligence to the edge in application domains such as computer vision and audio processing. Our work is positioned in this broad research field, then focusing on \ac{SED}.
To the best of our knowledge, this is the first attempt to use a student-teacher approach to perform sound event recognition directly on an IoT end-node in just 34.3 kB of RAM and 5.5 mW of power consumption.

\section{Knowledge Distillation}
\label{sec:knowledgedistillation}



\begin{figure}[!t]
\centering
\includegraphics[width=0.9\linewidth]{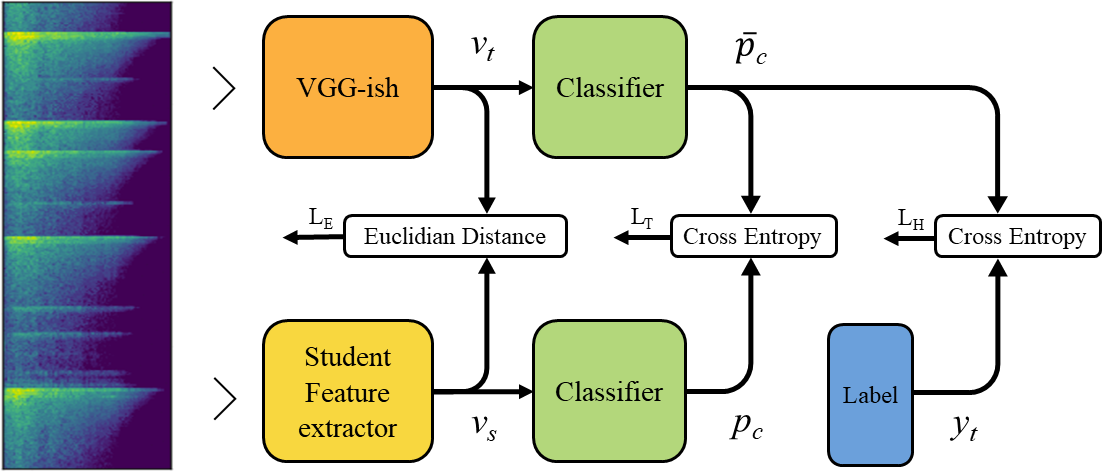}
 \caption{Block diagram of \ac{KD} using teacher intermediate features (L$_E$), soft teacher output (L$_T$) and dataset labels (L$_H$)}
\label{fig:student-teacher}
\vspace{-10pt}
\end{figure}

In this section, we give a brief overview of the Student-Teacher approach. A more detailed review is available in~\cite{cerutti2019IS}, where we present our proposed distillation strategy based on a compound loss function. In addition to this, here we introduce a two-stage distillation that provides a small but significant performance improvement.

Considering a generic architecture of a neural network, where the classifier follows a feature extractor, distillation can take place in different parts of the network. Figure~\ref{fig:student-teacher} graphically shows the idea behind the distillation process. The upper part represents the teacher network, the lower part is the student, and arrows indicate where the loss between the two networks is evaluated to train the student.

The original approach, \cite{hinton2015distilling}, replaces the hard labels with the teacher output in the soft loss:
\begin{equation}
\label{eq:soft}
\L_{\mbox{s}}(\X) = -\sum_{n=1}^N \sum_{c=1}^C \overline{p_c}(\x_n)\log(p_c(\x_n)), 
\end{equation}
where $\X=\left\{\x_1,\dots \x_N \right\}$ is the set of input features, $N$ is the number of samples, $C$ is the number of classes, $c=\left\{1,\dots,C\right\}$ is a generic class, $\overline{p_c}(\x_t)$ and $p_c(\x_t)$ are the logits of teacher and student respectively for input $\x_n$.

A further strategy, in line with~\cite{romero2014fitnets}, is to make the student learn how to replicate also the features produced by the teacher. The embedding loss $\L_{\mbox{e}}(\X)$ is thus defined as: 
\begin{equation}
\label{eq:emb}
\L_{\mbox{e}}(\X) = \sum_{n=1}^N  ||v_{t}(\x_n)-v_{s}(\x_n)||^{2},
\end{equation}
where $v_{t}(\cdot)$ and $v_{s}(\cdot)$ are the feature vectors produced by teacher and student networks respectively.

In~\cite{cerutti2019IS} we observed that the best solution is to combine  different losses via a linear combination:
\begin{equation}
\label{eq:allLoss}
\mathcal{L}(\X) =\alpha_{h} \mathcal{L}_{\mbox{h}}(\X) +\alpha_{s} \mathcal{L}_{\mbox{s}}(\X) + \alpha_{e} \mathcal{L}_{\mbox{e}}(\X),
\end{equation}
where $\L_{\mbox{h}}(\X)$ is the standard cross-entropy using the hard (one-hot) labels $\Y=\left\{\y_1,\dots \y_N \right\}$:
\begin{equation}
\label{eq:hardLoss}
\L_{\mbox{h}}(\X) = -\sum_{n=1}^N \sum_{c=1}^C \y_n\log(p_c(\x_n)),
\end{equation}


\subsection{Dataset}
\label{sec:experiments}
For \ac{SED} in outdoor urban environments, three datasets are often used in literature: \textit{UrbanSound8K}~\cite{salamon2014dataset}, AudioSet~\cite{audioset}, ESC50~\cite{piczak2015esc} and TUT Sound events 2017~\cite{Mesaros2016_EUSIPCO}. The latter is particularly attractive because it features real recordings and several comparative methods are available thanks to the related DCASE challenges~\cite{DCASE2017challenge}. Unfortunately, the task is very hard and the state-of-the-art accuracy is rather low. Moreover, the class distribution is highly unbalanced towards one class. Therefore the dataset does not allow a fair evaluation of \ac{KD} methods. ESC50 is also rather in line with our application scenario, but its size is relatively small (3 minutes of audio per class) and does not allow generalizing the results. Finally, we also discarded AudioSet because its video-based labels, referring to scenes instead of isolated sound-events, require consistent additional work to be aligned with the label required for our analysis. 
Therefore, we focused on \textit{UrbanSound8K}. It includes 8732 audio samples related to the city environment, with different sampling rates, number of channels and a maximum length of 4 seconds. Each recording has a unique label among 10 possible classes: air conditioner, car horn, children playing, dog bark, drilling, engine idling, gun shot, jackhammer, siren, and street music. 
All clips are taken from Freesound\footnote{"http://www.freesound.org"}, a vast collaborative database of audio samples. Following the recipe reported in \cite{salamon2014dataset}, we use 10-fold cross-validation and average scores.
However, we took one additional fold for validation: 8 folds are used as training data, one is used as validation and the remaining one for test (training-validation-test ratio is 0.8-0.1-0.1). Validation fold is one index less than the test one (for example, when the test fold is 9, the validation fold is 8). Performance is measured in terms of classification accuracy. In all experiments, the dataset is augmented through pitch shifting \cite{salamon2017deep}, with both positive (up) and negative (down) semitones with values {-2,-1,1,2}.

\begin{table}[t]
  \caption{Architecture of the models under analysis. "-" means that the layer is not active, "x" means that the layer is active.}
  \label{tab:models}
  \centering
\resizebox{\columnwidth}{!}{%
\begin{tabular}{lccccc}
Layer & VGGish/M$_{70M}$ & M$_{20M}$  & M$_{2M}$ & M$_{200k}$ & M$_{20k}$\\ \hline
Conv1 & 64      & 64 & 32  & 8    & 4   \\
Pool1 & x       & x  & x   & x    & x   \\
Conv2 & 128     &128 & 64  & 16   & 8   \\
Pool2 & x       & x  & x   & x    & x   \\
Conv3 & 256     &256 & 128 & 32   & 16  \\
Conv4 & 256     & -  & -   & -    & -   \\
Pool3 & x       & x   & x & x    & x    \\
Conv5 & 512     &256  & 128 & 64   & 16 \\
Conv6 & 512     & -     & -   & -    & -\\
Pool4 & x       & x & x   & x    & x   \\
Conv7 & -       & - & -   & 64   & 32  \\
Pool5 & -       & - & -   & x    & x   \\
FC1   & 4096    & 2048 & 512 & 256  & 64\\
FC2   & 4096    & 2048 & -   & -    & - \\
FC3   & 128     & 128 & 128 & 128  & 128\\
BatchNorm & x   & x & x  &  x   &  x    \\
GRU   & 20      & 20 & 20  & 20   & 20  \\
FC4   & 10      & 10 & 10  & 10   & 10 \\
\hline
\#Param & \ti72.1M & \ti18.0M & \ti1.88M & \ti202k & \ti30.6k \\
\end{tabular}
}
\end{table}
\subsection{Teacher}
State-of-the-art solutions for \ac{SED} are mostly \ac{CNN} fed with mel-spectrogram \cite{piczak2015environmental}\cite{zhang2018deep}\cite{zhang2017dilated}. However, to fully exploit the potential of the~\ac{KD} approach, rather than training from scratch our big \ac{CNN}, we employed the publicly available VGGish\footnote{"https://github.com/tensorflow/models/tree/master/research/audioset"} feature extractor~\cite{hershey2017cnn}, followed by a classification stage tailored on \textit{UrbanSound8K}. VGGish is trained on the Youtube-8M Dataset \cite{gemmeke2017audio} and it is expected to generalize well to other application contexts. Note that this fact introduces a further novelty in our work since distillation is performed on a different dataset than that used in the original training. 
VGGish converts \SI{960}{\milli\second} audio input mel spectrogram into a 128 dimensional embedding that can be used as input for a further classification model. The classifier can be shallow as the VGGish embeddings are more semantically representative than raw audio features. The VGGish architecture is described in Table \ref{tab:models}. For all the convolutional layers the kernel size is 3, the stride is 1 and the activation function is ReLu. Max pooling layers are implemented with a 2x2 kernel and a stride of 2.

The classifier consists of a \ac{GRU} followed by a fully connected layers and maps the VGGish embeddings into the 10 classes of \textit{UrbanSound8K}. A BatchNormalization layer is inserted between the feature extractor and the classifier to accelerate training. 

VGGish expects as input \SI{960}{\milli\second} of audio signal sampled at \SI{16}{\kilo\hertz}. Each clip in \textit{UrbanSound8K} is resampled and padded or cropped to get a length of 960*4=\SI{3840}{\milli\second}. The resulting signal is divided into 4 non-overlapping \SI{960}{\milli\second} frames. For each frame, the short-time Fourier transform is computed on \SI{25}{\milli\second} windows every \SI{10}{\milli\second}. The resulting spectrogram is integrated into 64 mel-spaced frequency bins, covering the range 125-7500 Hz, and log-transformed. This gives 4 patches of 96 x 64 bins that form the input of the VGGish.

\begin{figure}[t]
  \centering
  \includegraphics[width=0.8\columnwidth]{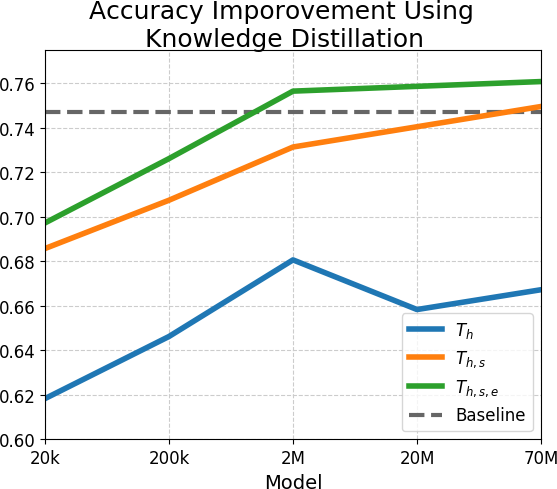}
  \caption{Average accuracy for the 4 different models in table \ref{tab:models} plus model M$_{70M}$. Each line represents a different training strategy. $T_h$ is standard training, using just hard labels. $T_{h,s}$ uses both hard and soft labels. $T_{h,s,e}$ minimize also the feature vector produced by an intermediate level. Dashed line is the teacher accuracy, here referred as baseline.}
  \label{fig:accuracyVsTraining}
\vspace{-10pt}
\end{figure}

\subsection{Distillation}
In this section, we analyze how the compression factor impacts on the final classification accuracy, using the standard loss and the loss compound described above.
Although our goal is to fit the classifier on an \ac{IoT} device, we consider four different degrees of compression, to better assess the potential and limits of the proposed approach, as reported in Table~\ref{tab:models}.
An heuristic adjustment of the layers and the number of filters is used to design the reduced networks. The model subscript M$_{x}$ approximately represents the number of parameters of the upstream part. 
Note that the student networks drastically reduce the feature extractor only. The classifier is re-trained, but it keeps the same architecture, i.e. a 20-unit recurrent layer followed by a 10-unit fully connected layers.


Figure \ref{fig:accuracyVsTraining} reports the results in terms of classification accuracy for the 4 models and when training from scratch on {\it UrbanSound8K} and when distilling from the VGGish teacher. In addition, we consider the model M$_{70M}$, which is a replica of the VGGish's architecture. All models considerably improve thanks to knowledge distillation with respect to being trained from scratch. Note also how the large models (M$_{20M}$, M$_{70M}$) perform much worse than VGGish and slightly worse than M$_{2M}$ when trained from scratch: this is due to the fact that {\it UrbanSound8K} is not large enough for such huge networks.  
Finally, it is worth noting that the VGGish baseline is outperformed also by all models with more than 2M parameters. This is mostly due to the domain adaptation that we are implicitly applying when distilling knowledge from the teacher. As a matter of fact, the student feature extractor is tailored to the new in-domain data from {\it UrbanSound8K}, providing an improvement over the more general purpose VGGish feature extractor trained on Youtube8-M. 

\begin{figure}[t]
  \centering
  \includegraphics[width=\columnwidth]{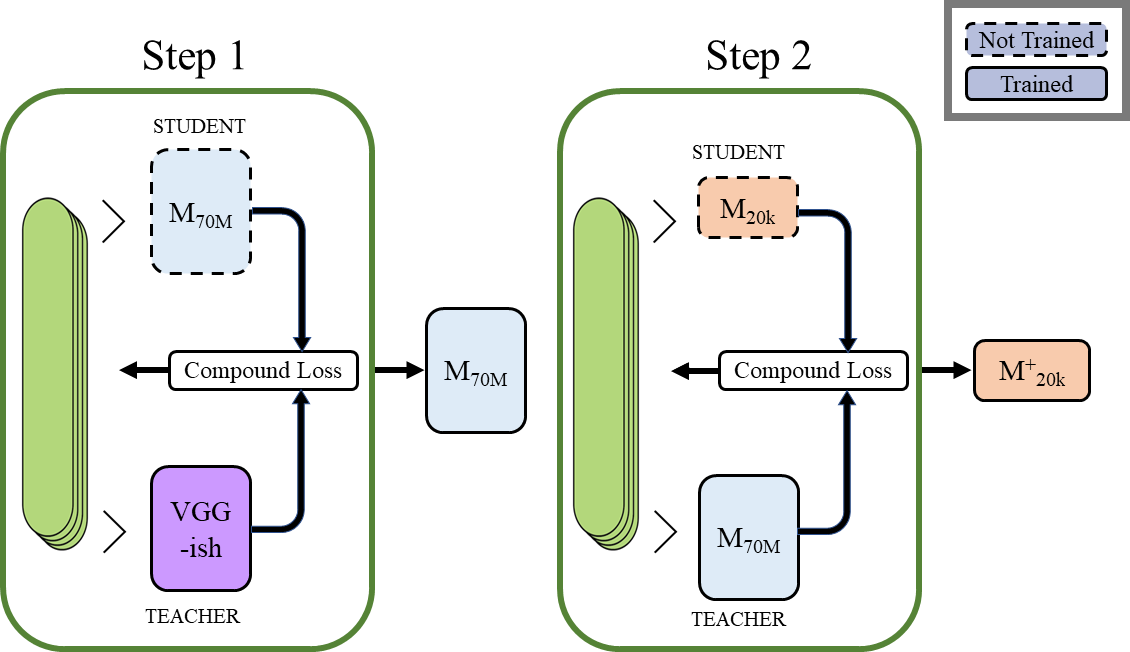}
  \caption{Two stage distillation: M$_{70M}$ is firstly trained from VGGish and the student M$_{20k}$ is distilled from M$_{70M}$ }
  \label{fig:two_steps}
  \vspace{-5pt}
\end{figure}

\subsection{Two-stage distillation}
The previous analysis, showing that we are jointly performing domain adaptation and parameter reduction, suggested us to investigate whether a further improvement can be achieved by separating these two processes.

Figure \ref{fig:two_steps} describes our proposed 2-stage distillation, where we train the smallest network using the M$_{70M}$ network as teacher. Note that M$_{70M}$ is still trained using the compound loss in Eq.~\ref{eq:allLoss}. Results are reported in Table~\ref{tab:2stage_performance} in comparison against training from scratch and using distillation from VGGish. The proposed approach achieves a 72.67\% accuracy on the test set, providing a 3 points improvement over a more traditional distillation strategy (69.7\%). Interestingly, the new M$_{20k}$ model is just 2 points below the VGGish baseline and 4 points below M$_{70M}$, in spite of using less than 0.0424\% of the parameter and 0.12\% of the operation. 

\begin{table}[ht]
  \caption{Accuracy of the M$_{20k}$ for: training from scratch, distillation from VGG-ish and distillation from M$_{70M}$}
  \label{tab:2stage_performance}
  \centering
\begin{tabular}{l|c|c|c}
 & Standard Training & From VGG-ish & From M$_{70M}$ \\ \hline
M$_{20k}$ & 61.83 & 69.72 & 72.67 \\ \hline
\end{tabular}
\end{table}


\section{Hardware resources and network requirements}\label{sec:hardware_analisys}
\subsection{Approximate hardware requirements per model}
 One interesting aspect towards tayloring our distilled networks for a resource constrained platform is to understand what are the computational and hardware requirements for a given model, or to understand what accuracy would be achievable with a given platform.
\begin{table}[ht]
  \caption{Approximate computational and memory requirements for each network}
  \label{tab:models_costs}
  \centering
\resizebox{\columnwidth}{!}{%
\begin{tabular}{lccccc}
  & VGGish & M$_{20M}$  & M$_{2M}$ & M$_{200k}$ & M$_{20k}$\\ \hline
\#Param & \ti72.1M & \ti18.0M & \ti1.88M & \ti202k & \ti30.6k \\
\#Operations & \ti1.72G & \ti608M & \ti148M& \ti13.6M & \ti2.11M \\
\#Buffer [B] & \ti614k & \ti602k & \ti301k & \ti76.0k & \ti 34.3k 
\end{tabular}
}
\end{table}

Table~\ref{tab:models_costs} reports an approximation of the computational requirements of each network: number of parameters to store, number of operations and buffer sizes. For memory requirements, we refer to an implementation using 8-bit quantized weights and buffers as specified in the CMSIS-NN library.

With an 8-bit quantization, each parameter takes one byte; therefore, the {\bf non-volatile memory} in bytes needed by a model equals the total number of weights and biases (first row of Table~\ref{tab:models_costs}).

Buffers keep the outputs of each layer available during propagation and are stored in the {\bf RAM}. The size of these buffers depends on the output dimensionality of each layer. However, the total amount of run-time memory required depends heavily on how efficiently buffers are implemented. The third row of Table~\ref{tab:models_costs} reports the RAM requirements of each model given the buffer design described in section \ref{sec:embeddedprogramming}.

The {\bf number of operations} (both multiplications and sums) depends on the layer type: in convolutional and maxpooling layers, each output pixel comes from a filter application.  Given kernel size $k$ and number of input channels $c$, each filter requires the following operations:
\begin{equation}
Ops_{filter} = 2\cdot c\cdot k^2.
\end{equation}
Therefore, the number of operations for each convolutional and maxpooling layer is:
\begin{equation}
Ops_{Conv} = Ops_{filter} \cdot out_{H} \cdot out_{W} \cdot out_{C},
\end{equation}

where $out_{H},\ out_{W},\ $and$ out_{C}$ are the output height, width an channel respectively.

For dense and \ac{GRU} layers the number of operations is the number of matrix multiplications:
\begin{equation}
Ops_{DotProduct} = 2\cdot in_{shape} \cdot out_{shape}.
\end{equation}
Gated recurrent unit requires also three element-wise products.

\subsection{Selection of the device class}
Table \ref{tab:platformIntro} shows a non-exhaustive list of processing platforms potentially adequate to be integrated into an end-device, with their power consumption and memory. In this section, we provide a qualitative analysis of what devices would be able to run the distilled models presented in the previous sections. This analysis is inevitably rough as many figures are approximated and the actual values cannot be derived analytically. Additionally, we are not considering the time and resources required for other processing stages (e.g., Mel-bins extraction); thus, the constraints are rather relaxed with respect to the actual application. Nevertheless, the devices in Table~\ref{tab:platformIntro} show substantial differences in~\ac{MIPS} and memory (in the order of powers of 10); therefore, the results of this study are still valid as long as we refer to classes of devices.

\begin{table}[th!]
 \caption{Examples of embedded platforms and their hardware capabilities.}
\resizebox{\columnwidth}{!}{%
\begin{tabular}{l|llll}
Board Name              & Flash{[}KB{]} & RAM{[}KB{]} & Power {[}mW{]} & MIPS\\ \hline
Arduino                 & 32            & 2           & 60 &  20          \\
ChipKit uc32            & 512           & 32          & 181 & 124.8           \\
{\bf STM32L476RG}      & 1024          & 128         & 26 & 80          \\ 
{\bf TI MSP432P4111} & 2048          & 256         & 23   & 58.56       \\ 
BeagleBone Black        & Ext           & 524288      & 2300 & 1607           \\
Raspberry Pi 3 B+ & Ext           & 1048576     & 5500  & 2800        
\end{tabular}
}
\label{tab:platformIntro}
\end{table}

{\bf Computational Cost}. To ensure real time classification, the network must process each 960 ms audio frame (\ti1 second) before the next frame arrives. Thus, classification time must be shorter than 1 second. Unfortunately, converting exactly the~\ac{MOPS} required by each model in the~\ac{MIPS} available on a given device, as reported in the datasheet, is not feasible. As a matter of facts, the number of instructions required by an operation depends on many factors, the most important being the actual implementation. In this analysis, we rely on the assumption that, on average, the number of operations is equal to the number of instructions. Typically, instructions are more than operations because each operation involves branch, load and store instructions. However, most of 32-bit microcontrollers support \ac{SIMD}, which allows up to four instructions in one clock cycle. We assume that these two effects are balanced and the number of operations is equal to the number of instructions. In the next sections, we will demonstrate that the actual throughput (operation per instruction) is larger than 1, but it gets close enough (1.8) in some situations. Given this assumption, in combination with the fact that other processing stages are not being accounted for, the \ac{MIPS} available on the device must be larger (with some margin) than the~\ac{MOPS} needed in one forward propagation.

{\bf Memory constraints} are also relevant: the RAM on-board must contain the intermediate values and the non-volatile memory should contain all network parameters.
However, non-volatile memory is not the main limit in our examples: whenever the network does not fit in the flash memory it does not respect one of the others two parameters (RAM or \ac{MIPS}). 

\begin{figure}[ht]
  \centering
  \includegraphics[width=\linewidth]{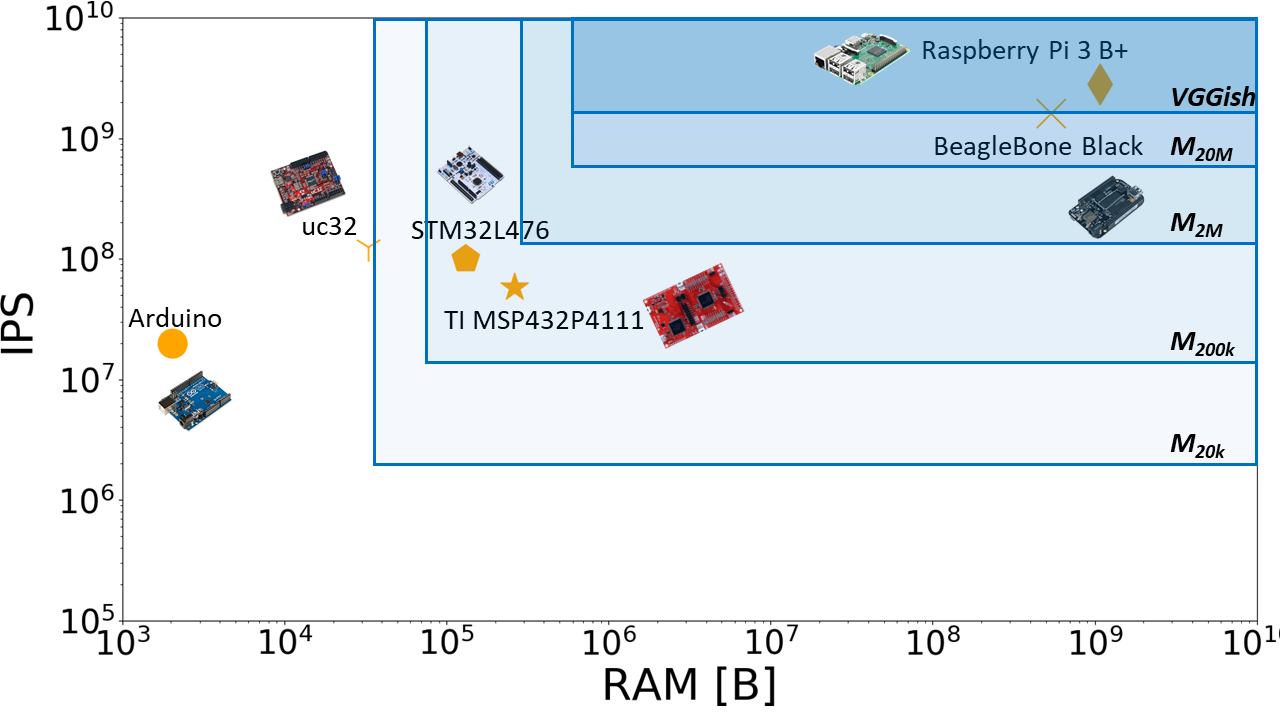}
  \caption{Models requirements vs hardware capabilities: a qualitative analysis Rectangles define network requirements : in terms of~\ac{MIPS} and RAM.}
  \label{fig:availablePlatform}
\end{figure}

Figure \ref{fig:availablePlatform} roughly compares the devices in Table \ref{tab:platformIntro} in terms of RAM and~\ac{MIPS} limitations against the compressed models each device can afford.
None of the proposed models fits in the Arduino platform due to its limited RAM. The last two platforms of Table~\ref{tab:platformIntro} (BeagleBone Black and Raspberry Pi 3 B+) have enough RAM and \ac{MIPS} to handle all models (actually VGGish fits only in Raspberrypi 3 B+). However, their power consumption is in the range of Watts that is not suitable for~\ac{IoT} applications. Therefore only devices ChipKit uc32, STM32L476RG, and TI MSP432P4111, which are approximately in the same class, are potentially suitable to host the distilled models. Uc32 is faster than both STM32 and TI platforms and would comfortably run the two smallest models in real-time, but its 32KB RAM is not large enough to fit even the \ti\SI{34.3}{\kilo\byte} needed for the buffers of the smallest model M$_{20k}$.
Concluding, M$_{20k}$ and M$_{200k}$ are the only suitable models for the target application and can be implemented in both the STM32L476RG and  TI MSP432P4111 platforms. In the next section, we detail our implementation of M$_{20k}$ in the STM32L476RG platform.

\section{embedded programming}
\label{sec:embeddedprogramming}
For the implementation of our \ac{SED} system, we selected a Cortex M4 architecture operating up to \SI{80}{\mega\hertz}, which is a good compromise between efficiency in processing and flexibility in power management. In particular, we worked on an STM3276RG Nucleo board as a development platform. Thus, the reference values for current consumption are \SI{420}{\nano\ampere} in standby mode and \SI{100}{\micro\ampere/\mega\hertz} in full running mode. The Floating Point Unit (FPU) features single precision and implements a full set of optimized \ac{DSP} instructions for fixed-point operations.



Cortex-M4 provides \ac{SIMD} instructions that operate on 8 or 16-bit integers. They are powerful for processing data such as video or audio, when full 32-bit precision is not necessarily required. The \ac{SIMD} instructions allow 2 x 16 bit or 4 x 8 bit operations to be performed in parallel \cite{SIMD_white}. 

\subsection{Quantization}
Feed-forward propagation through a neural network requires vector/matrix/tensor multiplication and convolution. Therefore, all the core features employed in signal processing can be used for neural network computations. For this reason, ARM developed the CMSIS-NN framework for neural network propagation on top of \ac{DSP} libraries \cite{lai2018cmsis}. The CMSIS-NN library maximizes performance and energy efficiency of common deep learning kernels on top of Cortex-M series cores.

Like for \ac{DSP}, truncation of floating point numbers to 8 or 16-bit fixed point numbers improves the execution time and reduces the memory footprint. According to \cite{lai2018cmsis}, 8-bit quantization achieves 4.6X improvement in runtime/throughput and 4.9X improvement in energy efficiency in image classification with CIFAR10 dataset.
On the other hand, since quantization implies loss of precision, we could expect a direct impact on the final prediction performance. However, the authors of \cite{lin2016fixed} experimented different kinds of quantization in image classification, achieving a 20\% drop in model size without significant loss of accuracy.  %
In the following, we describe the quantization process from a 32-bit floating-point to an 8-bit fixed-point representation. From now on, we will assume that floating point numbers has infinite precision and we will use the nomenclature typically used for \ac{DSP}. 

Quantization describes a real number with finite resolution. When a uniform quantization is applied, three parameters are used to define the fixed point representation: bit-width, step-size (resolution) and dynamic range \cite{lin2016fixed}. These parameters are correlated by the following expression:
\begin{equation}\label{eq:range}
Range = Stepsize \cdot 2^{bitwidth - 1},
\end{equation}
where $bitwidth - 1$ accounts for the bit used to represent the sign. $Stepsize$ is the minimum step between two fixed point numbers and will be always chosen as a power of two for convenience with binary representation.  

An equivalent formulation of Eq. \ref{eq:range} can be obtained by considering the number of bits used for the decimal and integer parts of numbers: 
\begin{align*}
    Range &= 2^{integer} \\
    Stepsize &= 2^{-decimal} \\
    bitwidth &= integer + decimal + 1 
\end{align*}
where, $integer$ is the number of bits used to represent the integer and $decimal$ is the number of bits used for the decimal part. +1 in the last equation accounts for the sign bit. Basically, given a fixed number $bitwidth$ of bits, increasing the range by increasing the number of bits for the integer part degrades the resolution. Conversely, decreasing $Stepsize$, by allocating more bits to the decimal part, leads to a range reduction.

The quantization error $e_q$ is the difference between the infinite precision number and the quantized representation:
\begin{equation}
    e_q = x - Q(x)
\end{equation}
where $x$ and $Q(x)$ are the input and the quantized output.
If we treat the input as a random variable with a probability density function $f_{x}$ (see \cite{sun2008image}), the mean square quantization error is the combination of two different errors, namely granular error ($\mathrm{MSE}_{q,g}$) and overload error ($\mathrm{MSE}_{q,o})$): 
\begin{align*}
    \mathrm{MSE}_{q,g}&=\sum_{i=1}^{\mathrm{N}} \int_{d_{i}}^{d_{i+1}}(x-Q(x))^{2} f_{x}(x) \mathrm{d} x \\
    \mathrm{MSE}_{q,o}&=\int_{-\infty}^{d_{0}}(x-Q(x))^{2} f_{x}(x) \mathrm{d} x\ + \\ &+\int_{-\infty}^{d_{N}}(x-Q(x))^{2} f_{x}(x) \mathrm{d} x
\end{align*}
where $N$ is the number of quantization levels (in our case $2^{(bitwidth)}$) and $d_i$ are the decision levels, i.e. any number between $d_i$ and $ d_{i-1}$ is coded with the same fixed-point representation, usually $(d_i + d_{i-1}) / 2$. 
The two mean square errors are related to $Stepsize$ and $Range$ respectively. If $Stepsize$ is reduced then MSE$_{q,g}$ decreases; on the other hand if $Range$ is made wider, MSE$_{q,o}$ decreases. 

The figure of merit linked with \ac{MSE} is the \ac{SQNR}, defined as:
\begin{equation}
    \mbox{SQNR} = \frac{E\left[x^{2}\right]}{E\left[e_q^{2}\right]} = \frac{E\left[x^{2}\right]}{\mathrm{MSE}_q}
\end{equation}

Therefore, the goal is to define the optimum trade-off between $Range$ and $Stepsize$ to minimize the \ac{SQNR} in the target application. In the following, we describe two different strategies to perform quantization using the quantization parameters and its figure of merit described above.

\subsection{Quantization Design} 
\label{sec:QuantizationDesign}
Applying quantization to neural networks has different requirements with respect to other contexts, as image quantization or audio quantization. In particular, we are not interested in preserving an accurate representation of all activation outputs or weights for each layer. Conversely, we want that the final prediction is as close as possible to the prediction of the network with a 32-bit floating-point representation. To summarize, accuracy is the most relevant metric. 

An exhaustive search of all the possible $integer$/$decimal$ combinations, evaluating the final accuracy, is not feasible in practice. In the smallest network, i.e. M$_{20k}$, the number of $integer$/$decimal$ digits would be 28, and each of them can vary between 0 and 7 (setting the decimal digit decides the integer part when $bitwidth$ is fixed). 
Therefore, the number of possible combinations is given by the permutation with repetition: $7^{28} = 4.5 * 10^{23}$. 

The two solutions presented in this paper target the maximization of the \ac{SQNR} or the reduction of the overload error.

The first approach to compute \ac{SQNR} applies quantization on each variable (weights, activation) one at a time. For the weights, we choose the number of decimals that maximizes the \ac{SQNR}. For the intermediate outputs 
we compute the \ac{SQNR} running a forward propagation in floating-point on the training dataset and testing all the possible integer/decimal values (in the range from 0 to $bitwidth$). 
Each activation output is analyzed independently, eventually leading to different quantization trade-offs. 

This approach relies on \cite{lin2016fixed}, where the overall \ac{SQNR}, here defined as $\gamma$, is the harmonic mean of the \ac{SQNR} of all preceding quantization steps:
\begin{equation}
\frac{1}{\gamma_{\text {output}}}=\frac{1}{\gamma_{a^{(0)}}}+\frac{1}{\gamma_{w^{(1)}}}+\frac{1}{\gamma_{a^{(1)}}}+\cdots+\frac{1}{\gamma_{w^{(L)}}}+\frac{1}{\gamma_{a^{(L)}}}
\end{equation}
It turns out that maximizing the single \ac{SQNR}s maximizes the overall \ac{SQNR}, regardless of where quantization happens (at the first layers or at the last layers). 

The second approach is based on the different effects of granular and overload errors.
It is reasonable that a single number with a high quantization error due to overload will affect the overall forwarding of the neural network. Conversely, the ensemble of small granular quantization error may not change the argmax of softmax layer, that is the value used to determine prediction and accuracy. Therefore, we select the integer/decimal ratio that reduces the probability of having values out of the quantization range. In particular, we set the number of bits for the integer part $i$ such as
\begin{equation}
    \min(i) : P(|x| < 2 ^{i}) < P_{th}
\end{equation}
where $x$ is the floating-point input and $P_{th}$ is the probability of overload that one is willing to accept. 
This approach differs from the previous one for three reasons. (i) Numbers with large overload error have $bitwidth$ of the same importance than numbers with small overload error. (ii) Thus, each overload is considered uniformly, without accounting for the granular error but taking into only the overload error. (iii) This second approach has a hyper-parameter, i.e. the probability threshold, which requires a fine tuning on the training set. We heuristically set the threshold to $10{^{-4}}$.

We estimate the probability density function $f(x)$ using the whole training set. The percentage of values out of the range is counted for all the possible decimal values involved in the network feed-forward propagation.

\begin{figure}[t]
  \centering
  \includegraphics[width=\linewidth]{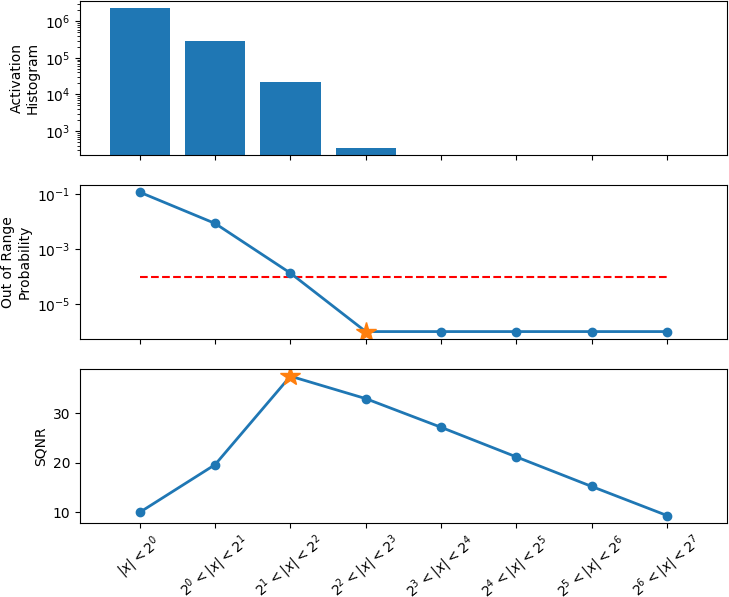}
  \caption{The upper part shows a histogram of the activation values of one layer,  between consecutive powers of 2. The lower part shows an example of the behaviour of the two quantization schemes investigated here. The central graph concludes that 3 decimals should be used for the integer part, because it is the first point in which probability goes below the threshold. The bottom line shows that two digits for the integer part are the solution with maximum \ac{SQNR} and therefore the minimum \ac{MSE}}
  \label{fig:hist}
\vspace{-8pt}
\end{figure}

The upper part of Figure~\ref{fig:availablePlatform} is the histogram of the absolute value of the activation outputs between two consecutive powers of two. Most of the values (more than $10^6$) are between -1 and 1 (or equivalently $|x| < 2^0$). Oppositely, just a small set of numbers (around $10^2$) are such that $2^2 < |x| < 2^3$; they are depicted in the fourth bar.
The central part of Figure~\ref{fig:hist} depicts the relative distrbution for each class of values, e.g. the estimated probability that a value is inside a given range related to two consecutive powers of 2. The values confirm what bars show: values in $|x| < 2^0$ are the most likely and values in $2^2 < |x| < 2^3$ appears only with a probability of $10^{-5}$.
Finally, the plot at the bottom depicts the \ac{SQNR} using the number of bits for the decimal part available for a given range in the x axis (the range determines the bits for the integer part). For example, the most left point is the \ac{SQNR} using 0 bits for integer part (which allows describing numbers in the range $0\leq|x| < 2^0$) and 7 bits for the decimal part. 
In this particular case, corresponding to a layer of the network, the optimum number of bits for the integer part obtained by maximizing the \ac{SQNR} (three) does not correspond to what obtained by setting the probability threshold to $10^{-4}$ (four). 

\subsection{Firmware Programming}

To transfer the neural model from to CMSIS-NN we developed an automatic python script to export the Keras model in an header file containing the weights to use in the firmware. To do so, firstly, we reordered the weights following the convention of both source and destination framework, then we quantized the weights using both quantization schemes. In this work, we set $bitwidth$ to 8 since we privilege execution time and power consumption with respect to accuracy. Note that we implemented a slightly different version of the M$_{20k}$ model, replacing the \ac{GRU} layer with a vanilla \ac{RNN}. The reason is that the implementation of a GRU layer in CMSIS-NN requires an additional effort while the accuracy does not change significantly. 

The CMSIS-NN functions require different arguments: input dimension, number of channel and so on. These parameters should be extracted from the neural network and loaded in the microcontroller. Some of them requires extra processing, like shifting of weights and bias in internal operations. A detailed description of these parameters follows here.

Suppose that we are in an intermediate layer $i$ and that the quantization procedure concludes that the number of integer bits $m$ and the number of decimal bits $n$ for weights, bias, input and output are $m_w$, $n_w$, $m_w$, $m_b$, $m_i$, $n_i$, $m_o$, $m_o$.
The neural network operations include always a linear combination of weights and input, so that 
\begin{equation*}
     output = input * weight + bias.
\end{equation*}
In multiplications between fixed point numbers, the number of decimal digits of the result are given by the sum of the number of decimal digits of the two operands. To sum the bias to this intermediate value, we need to use the same number of decimals. Thus, we shift the bias to make it match. In most of the cases it is a left shift. Finally, the output must have a certain number of decimals to get back in a fixed bitwidth format, so we apply a further shift. This last shift is opposite to the previous one and in CMSIS-NN is referred to as right shift. 
These concepts are expressed in formulas, implemented in the header files by means of a macro:
\begin{align*}
    left\_shift &= (n_i + n_w) - n_b \\
    right\_shift&= (n_i + n_w) - n_o
\end{align*}

Another parameter to bear in mind is the size of the buffers. To necessarily instantiate buffers in the program, they need to be known in advance. During inference, intermediate values are discarded, and only the final prediction and the network state (in case of recurrent neural network) are kept. As a consequence, it is possible to use the same buffers in multiple layers. 

\begin{table}[t]
\caption{buffers size required for each layer. Reuse of buffers allows memory saving. Two buffers (A,B) must contain maximum value in odd index and even index of size$_{output}$ respectively. }
\label{tab:buffers}
\centering
\begin{tabular}{l|llllll}
                 & I    & C1    & C2    & C3   & C4  & C5  \\ \hline
size$_{output}$ & 6144 & \textbf{23312} & \textbf{10440} & 4368 & 720 & ... \\
buffer           & A    & \textbf{B}     & \textbf{A}     & B    & A   & B  
\end{tabular}
\vspace{-15pt}
\end{table}

Each layer needs a pair of buffers able to contain output and input. To find the minimum size for these two buffers, we create a vector with the number of element between layers, $size_{output} = a_0, a_1, a_2, ...$, where $a_0$ is the input size. For each layer, the 2 buffers will switch their role: in the first convolutional layer, the input will be $a_0$ and the output $a_1$. For the next layer, $a_1$ is the input, $a_2$ is the output and so on.
Table \ref{tab:buffers} shows the sequence of input and output for each layer in the implemented network. It shows that the two buffer sizes should be selected accounting for the maximum in odd and even indexes in the vector.

Finally, the CMSIS-NN framework requires also an additional small buffer for intermediate calculations, which can be reused in the implementation too. 

\section{Results}
We evaluate the porting of our \ac{SED} model to the microcontroller, in terms of power consumption, execution and recognition accuracy.

\subsection{Accuracy}

\begin{figure}[t]
  \centering
  \includegraphics[width=\linewidth]{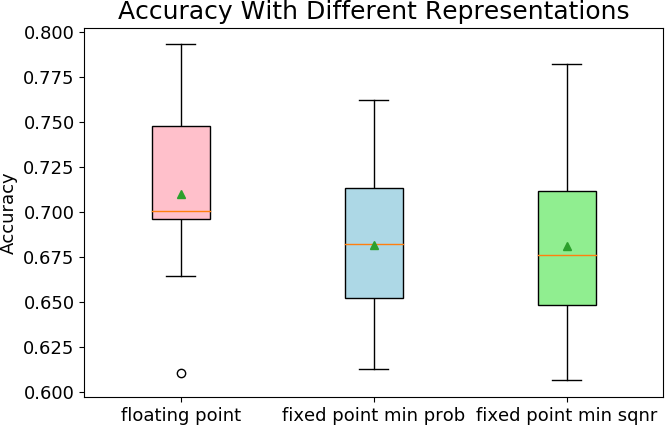}
  \caption{Accuracy over the 10 folds for \textit{Urbansound8K} dataset. Pink: floating point implementation. Blue: quantization by maximizing the \ac{SQNR}. Green: quantization by minimizing the probability of overload error}
  \label{fig:boxplot}
\vspace{-2pt}
\end{figure}

Input data from the test-set are sent by UART to the \ac{MCU} using a Python script. The forward propagation is computed inside the microcontroller that provides the prediction results on the same bus for accuracy evaluation. 
Following the recipe in \cite{salamon2014dataset}, we used a 10-fold cross validation and average scores. For each test fold, we load models in the microcontroller with the quantization parameters computed on the related training set.
Figure \ref{fig:boxplot} depicts the accuracy over the 10 folds. The average accuracy shows a decrease of performance in both quantization schemes, but it is limited to a 2\% drop overall if compared to the floating-point version (pink). Note that this minor performance drop could be limited by fine-tuning the quantized network~\cite{Zeng-2019}.

\begin{figure}[t]
  \centering
  \includegraphics[width=\linewidth]{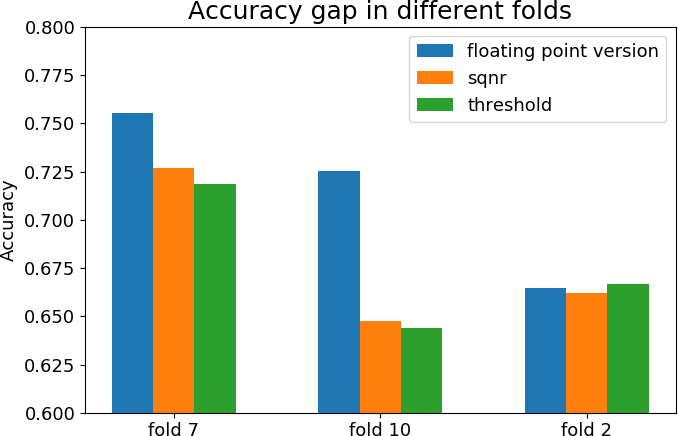}
  \caption{Accuracy gap between floating point network and quantized versions for different folds. Accuracy degradation in fold 10 is more evident than in fold 2}
  \label{fig:bargraph}
\vspace{-8pt}
\end{figure}

Looking into more details, this performance deterioration is strongly dependent on the train/test-set split of each fold. Figure \ref{fig:bargraph} reports the performance for three folds (train/test configurations): fold 7, fold 10, fold 2. When testing on fold 7, the performance deterioration is in line with the average accuracy decreases of around 2\%. On the other hand, when fold 10 is selected as a test set, the accuracy drop is more consistent (8\%). Finally, the impact of quantization error becomes negligible, or it even improves the classification results, in fold 2.
This behaviour is related to the robustness of the original floating point model.
As pointed out by Piczak \cite{piczak2015environmental}, some classes are more difficult to detect by a convolutional neural network, because of their short scale temporal structure (drilling, engine idling, jackhammer). Whenever the floating point network correctly classifies these classes, the difference between the likelihoods is probably shallow and the errors introduced by quantization can lead to a misclassification, leading to huge performance degradation. Similarly, the gap between the floating point and quantized network decreases or disappear when the reference network already does not classify correctly the difficult classes, as performance is already bad. 

\begin{figure}[t]
  \centering
  \includegraphics[width=0.85\linewidth]{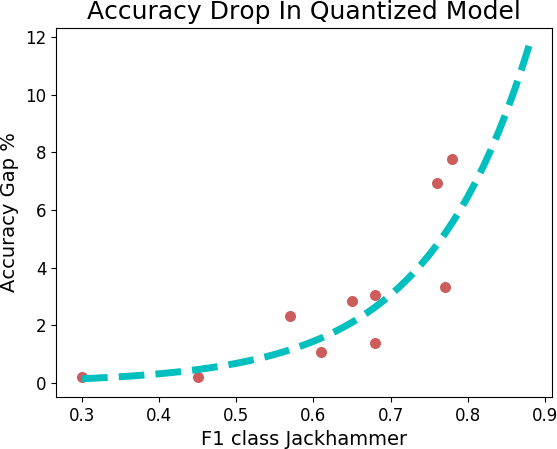}
  \caption{Relation between performance in a specific class (Jackhammer) and the drop in accuracy due to quantization error. Due to its short-time duration, the class is difficult to detect for the convolutional model, thus its classification gets harder in the quantized model}
  \label{fig:jackhammer}
  
\end{figure}

To confirm that, we compared the accuracy gap due to quantization and the F1 for these problematic classes. In all three cases, we confirmed that whenever the F1 score is high the accuracy drop increases, following an exponential profile as Figure \ref{fig:jackhammer} shows. The figure refers to the class "Jackhammer".

\subsection{Execution Time and Power Consumption}
In section \ref{sec:experiments}, we estimated the execution time of several network architectures by assuming that the number of operations is equivalent to the number of instructions, this way allowing a comparison between different platforms in terms of \ac{MIPS} available from the datasheets. In this section, we evaluate the actual execution time and the power consumption of our implementation of the smallest model M$_{20}$. 
Measurements are done on the real prototype using the same system used to compute the accuracy. We measure the performance layer by layer to understand how each component contributes to the overall processing time. Table~\ref{tab:layer_by_layer} reports the evaluation results. 

\begin{table}[ht]
\vspace{-9pt}
  \caption{Execution time on the real platform, layer by layer. In the bottom, overall execution time using CMSIS-NN and without DSP optimization (Plain C)}
  \label{tab:layer_by_layer}
  \centering
\resizebox{\columnwidth}{!}{
\begin{tabular}{lccccc}
Layer & Output Channel & \#kop & Exec. Time [ms] & MOPS\\ \hline
Input$_{96,64}$ & 1 & - & - & - \\
Conv1 & 4 & 419.61 &  63.93 & 6.56\\
Pool1 & x & 23.31 & 10.17 & 2.29 \\
Conv2 & 8 & 751.68 & 20.13 & 37.34\\
Pool2 & x & 24.84 & 8.32 & 2.99\\
Conv3 & 16 & 628.99 & 12.90 & 48.76\\
Pool3 & x & 11.09 & 3.63 & 3.03\\
Conv4 & 16 & 207.36 & 3.88 & 53.44 \\
Pool4 & x & 2.16 & 0.69 & 3.15\\
Conv5 & 32 & 27.65 & 0.60 & 46.23\\
Pool5 & x & 0.58 & 0.17 & 3.31\\
FC1   & 64 & 8.19 & 0.22 & 38,10\\
FC2   & 128 & 16.38 & 0.42 & 38.73\\
RNN   & 60 & 22.56 & 0.55 & 40,58\\
FC3   & 10 & 1.20 & 0.04 & 33.33\\
\hline
Total & - & 2145.61 & \textbf{125.6} & 17.08 \\  
Plain C & - & 2145.61 & 291.4 & 7.36
\end{tabular}
}
\vspace{-8pt}
\end{table}


The CMSIS-NN framework implements a "basic" and a "fast" version of convolutional layers. The latter uses assembly directives to speed-up execution, especially by means of \ac{MAC} and \ac{SIMD}. The only constraint for using this faster implementation is that the number of channels must be a multiple of 4 for 8-bit fixed-point quantization. This is the reason why, the number of channels of the convolutional layers of all architectures described in Table~\ref{tab:models} are multiple of 4.
Problems arise on the first layer, whose input is the Mel spectrum that has just one channel. This does not allow us to use the optimized version for convolutional layers and for this reason the first convolutional layer takes more than half of overall execution time, with a throughput of 6.56 \ac{MOPS} (see line 2 of Table~\ref{tab:layer_by_layer}). The second convolutional layer implements the fast version and takes just \SI{20}{\milli\second}, even if it is more computationally intense, performing 37.34 \ac{MOPS}. This highlights how important is the parallelization and how much a proper parallelization in the first layer can reduce drastically the overall execution time. 

To stress more the importance of an efficient framework for deep neural networks, we used a reference implementation of plain C (without explicit \ac{SIMD} directives) for convolutional, maxpooling and fully connected layers. The comparison is in the last two lines of Table \ref{tab:layer_by_layer}. The total execution time is speed-up of x2.32 with the fast implementation of CMSIS-NN with respect to plain C.

In Section \ref{sec:hardware_analisys}, we stated that the combination of x4 parallelization and overhead, due to branches and load-store instructions, makes the microcontroller able to perform an 8-bit operation in just one instruction. The selected platform executes 80 \ac{MIPS}, thus we expect 80 \ac{MOPS}, but the overall throughput is different in real-time measurements (average 17 \ac{MOPS}). Looking at the peaks in layer conv4, 53.44 \ac{MOPS} is still far from our estimation of 80 \ac{MOPS}. 
It means that parallelization does not fully compensate the overhead due to load-store and branches and we need approximately two instructions to execute an operation.
The average (17.08 \ac{MOPS}) is far from this peak level, mainly because of the first layer, that is the dominant part and it is not fully parallelized.

\section{Conclusions}
\label{sec:conclusions}
In this work, we described the whole process from a state-of-the-art model for sound event detection to its energy efficient implementation in a microcontroller, targeting \ac{IoT} applications. Firstly, we demonstrated that knowledge distillation can be effective also for extreme compression rates, achieving models suitable for real time applications on IoT nodes. Then, we introduced a two-step distillation to further improve the performance of the student network. 
Furthermore, we moved to the description of two quantization strategies, concluding that they perform in a similar way. Maximization of \ac{SQNR} is generally preferred with respect to the probabilistic approach, because it does not require any hyperparameter. Both 8-bit quantization schemes were applied to the smallest distilled model resulting in a 2 percent points drop in accuracy, in comparison with the original floating point version. 
The final implementation on the microcontroller has a propagation time of 125 ms for each 1-second-audio-clip using just 5.5 mW average power and 34.3 kB of RAM. We have shown that an efficient framework for neural network, like CMSIS-NN, speed-up significantly the execution.

One interesting extension of our work would be to combine the distillation and quantization step, using the soft label to train the quantized networks. Finally, in future works we will implement the whole chain on an \ac{IoT} node, including sensor acquisition, feature extraction and transmission of the classification outcome to the cloud. 



\bibliographystyle{IEEEtran}

\bibliography{mybib}
\end{document}